\documentclass{ws-p10x7}

\begin{document}

\title{Lepton Transmutations from a Rotating Mass Matrix}

\author{TSOU Sheung Tsun}

\address{Mathematical Institute, University of Oxford,\\
  24-29 St. Giles', Oxford OX1 3LB, United Kingdom\\E-mail: 
  tsou\,@\,maths.ox.ac.uk}

\twocolumn[\maketitle\abstract{
Fermion mass matrices generally rotate in generation space under scale 
changes, which can lead to fermions of different generations transmuting 
into one another.  The effect is examined in detail and its 
cross-section calculated for $\gamma + \ell_\alpha \longrightarrow 
\gamma + \ell_\beta$ with $\ell_\alpha \neq \ell_\beta$ the charged leptons 
$e, \mu$, or $\tau$.  For the (conventional) Standard Model, this 
is weak and probably undetectable, though with some notable exceptions.  
But for the Dualized Standard Model, which we advocate and have already 
used quite successfully to explain quark mixing and neutrino oscillation, 
the effect is larger and could be observable.  Estimates of
transmutational decays are also given.}]

\section{Introduction}
By a rotating mass matrix we mean one which undergoes unitary
transformations through scale changes, as a result of the
renormalization group equation, as can happen in many gauge
theories.  This means that even if the mass matrix $m$ is diagonal (in
generation space) at a certain scale, it will not remain so as the
scale changes.  Hence in general we can expect nonzero transition
between fermions of different generations:
\begin{equation}
\ell_\alpha \longrightarrow \ell_\beta, \quad \alpha \ne \beta,
\label{transmut}
\end{equation}
such as $e \to \mu,\ e \to \tau,\ \mu \to \tau$.  We shall use the
term `transmutation' for this direct transition to distinguish it from
e.g.\ $e \to \mu$ conversion via FCNC.

Here I shall concentrate on transmutational
processes\cite{impromat,photrans} in the standard model (SM) and, in greater
detail, in the dualized standard model (DSM).

\section{Mass matrix rotation}
In the SM, because the leptonic MNS mixing matrix\cite{mns} $U$ is
nontrivial\cite{expmix}, the mass matrix $L$ for the charged leptons
will rotate as a result of the following term in the linearized
RGE\cite{rge}: 
\begin{equation}
\frac{dL}{d\mu} = \frac{3}{128 \pi^2} \frac{1}{246^2} (ULU^\dag)
(ULU^\dag)^\dag L + \cdots,
\label{smrge}
\end{equation}
where $ULU^\dag=N$ the neutrino (Dirac) mass matrix.  Therefore $L$
cannot be diagonal at all scales.  The magnitude of the off-diagonal
elements will depend on poorly known or unknown quantities such as the
mixing $U$ and the Dirac mass $m_3$ of the heaviest neutrino.  If we
take the present popular theoretical biases, namely that $U$ is
bimaximal\cite{bimax} and that $m_3$ is around the $t$ quark mass,
then (\ref{smrge}) gives
\begin{eqnarray*}
\langle \mu | \tau \rangle & {\rm changes\ by} & \sim 5.5 \times
10^{-3}\ {\rm GeV} \\
\langle e | \tau \rangle & {\rm changes\ by} & \sim 1.8 \times 
10^{-7}\ {\rm GeV}\\
\langle e | \mu \rangle & {\rm changes\ by} & \sim 1.1 \times 
10^{-8}\ {\rm GeV}
\end{eqnarray*}
per decade change in energy.

\begin{figure}
\centering
\epsfxsize120pt
\includegraphics[angle=-90,scale=0.55]{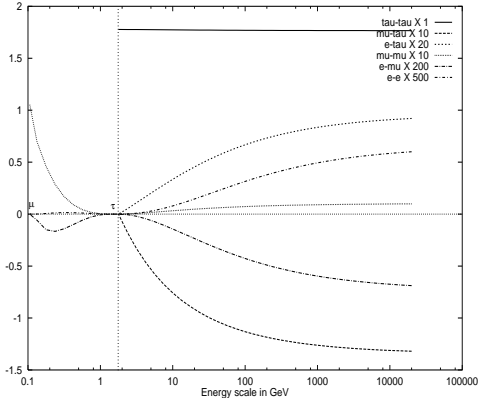}
\caption{Elements of the rotating mass matrix in GeV for charged 
leptons in the DSM scheme.}
\label{masmat}
\end{figure}

In the DSM\cite{dsm}, the fermion mass matrix is of the following
factorized form: 
\begin{equation}
m = m_T \left( \begin{array}{c} x \\ y \\ z \end{array} \right)
      (x, y, z),
\label{mmat}
\end{equation}
where $m_T$ is essentially the mass of the heaviest generation.  Under
renormalization $m$ remains factorized, but the vector $(x,y,z)$
changes as
\begin{equation}
\frac{d}{d\mu} \left( \begin{array}{c} x \\ y \\ z \end{array} \right)
   = \frac{5}{32 \pi^2} \rho^2 \left( \begin{array}{c} x_1 \\ y_1 \\ z_1
      \end{array} \right),
\label{dsmrge}
\end{equation}
where $\rho$ is a (fitted) constant and
\begin{equation}
x_1 = \frac{x(x^2-y^2)}{x^2+y^2} + \frac{x(x^2-z^2)}{x^2+z^2},
   \ \ \ {\rm cyclic}.
\end{equation}
The off-diagonal elements have been calculated explicitly, using 3
free parameters determined by fitting experimental mass and mixing
parameters (giving sensible predictions for the remaining
paramenters)\cite{phenodsm}. These are shown in Figure \ref{masmat}.  
Hence the
results we report below are entirely parameter-free.

\section{Lepton states}
To define lepton states one must diagonalize the mass matrix, but
since the eigenvectors depend on scale, there is no canonical recipe.

We suggest two quite different schemes for exploration: fixed
scale diagonalization (FSD) which is applicable to SM, and 
step-by-step diagonalization (SSD) which is an inherent part of 
DSM\cite{dsm,impromat,photrans}.

\section{Transmutational decays}
Using the results of \S2 we give estimates for the branching ratios of
transmutational decays\cite{impromat}.    With some exceptions, SM
(with FSD) estimates are all far below present experimental bounds and
are hence not so interesting.  One exception is the process $\mu^- \to
e^-e^+e^-$, where one could get a branching ratio of $10^{-3}$
(experimental limit $10^{-12}$), if one applied FSD naively.  This
shows how sensitive these calculations are to transmutational model
and/or diagonalization scheme.

The parameter-free estimates in DSM give branching ratios which
are in general larger but still below present experimental limits, as
indicated in Table \ref{decays}.  It is important to note that because
of SSD the branching ratios of transmutational leptonic decays are
automatically zero to first order (Figure \ref{masmat}).  
The $\pi^0$ decay is of particular
interest as being less than one order from the experimental bound.
\begin{table}
\caption{Branching ratios of transmuational decays.}\label{decays}
\begin{tabular}{|l|c|c|} 
 
\hline 
 
\raisebox{0pt}[12pt][6pt]{Decays} & 
 
\raisebox{0pt}[12pt][6pt]{DSM est.} & 
 
\raisebox{0pt}[12pt][6pt]{Expt limit} \\
 
\hline

\raisebox{0pt}[12pt][6pt]{$Z^0 \to \tau^- \mu^+$} & 
 
\raisebox{0pt}[12pt][6pt]{$4 \times 10^{-8}$} & 
 
\raisebox{0pt}[12pt][6pt]{$1.2 \times 10^{-5}$} \\

\raisebox{0pt}[12pt][6pt]{$\pi^0 \to \mu^- e^+$} & 
 
\raisebox{0pt}[12pt][6pt]{$3 \times 10^{-9}$} & 
 
\raisebox{0pt}[12pt][6pt]{$1.7 \times 10^{-8}$} \\

\raisebox{0pt}[12pt][6pt]{$\psi \to \mu^+ \tau^-$} & 
 
\raisebox{0pt}[12pt][6pt]{$6 \times 10^{-6}$} & 
 
\raisebox{0pt}[12pt][6pt]{not given} \\

\raisebox{0pt}[12pt][6pt]{$\Upsilon \to \mu^+ \tau^-$} & 
 
\raisebox{0pt}[12pt][6pt]{$2 \times 10^{-6}$} & 
 
\raisebox{0pt}[12pt][6pt]{not given} \\

\raisebox{0pt}[12pt][6pt]{$\mu^- \to e^- \gamma$} & 
 
\raisebox{0pt}[12pt][6pt]{0} & 
 
\raisebox{0pt}[12pt][6pt]{$4.9 \times 10^{-11}$} \\

\raisebox{0pt}[12pt][6pt]{$\mu^- \to e^- e^+ e^-$} & 
 
\raisebox{0pt}[12pt][6pt]{0} & 
 
\raisebox{0pt}[12pt][6pt]{$1.0 \times 10^{-12}$} \\\hline
\end{tabular}
\end{table}

\section{Photo-transmutation}
We studied\cite{photrans} (mainly for DSM) the following
\begin{equation}
\gamma +\ell_\alpha  \longrightarrow \gamma +\ell_\beta , \quad \alpha
\ne \beta.
\end{equation}

\begin{figure}
\centering
\epsfxsize120pt
\includegraphics[angle=-90,scale=0.3]{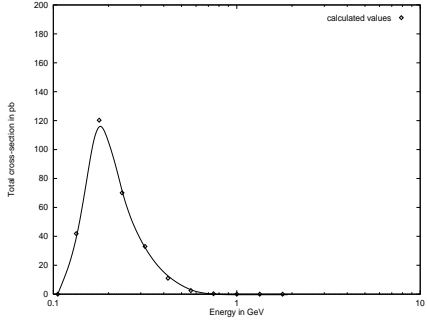}
\caption{Total cross section for $\gamma e \to \gamma \mu$.}
\label{total}
\end{figure}

We calculated the cross sections for: 
$\gamma e \to \gamma \mu,\ \gamma e \to \gamma \tau,\ \gamma \mu \to
\gamma \tau$, leaving out $\tau$-initiated reactions as being
experimentally unrealistic at present.  
\begin{figure}
\centering
\epsfxsize120pt
\includegraphics[angle=-90,scale=0.25]{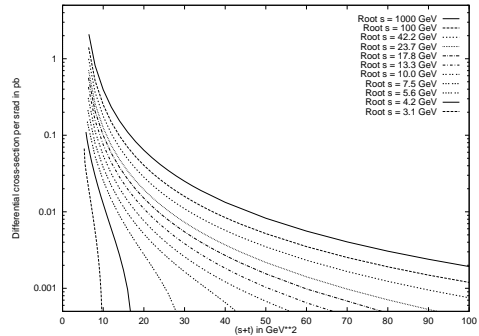}
\caption{Differential cross sections for $\gamma e \to \gamma \tau$.}
\label{cross}
\end{figure}
A sample of the DSM results is
presented in Figure \ref{cross},  for a range of c.m.\ energies $\sqrt{s}$.
Because of the calculated form of the rotation matrix, we get in
general:
$\gamma \mu \to \gamma \tau > \gamma e \to \gamma \tau > \gamma e \to
\gamma \mu.$
However, at low energies $\gamma e \to
\gamma \mu$ becomes quite sizeable, as seen in Figure \ref{total},
where the total cross section has a peak of $\sim$ 100 pb at c.m.\
energy $\sim$ 200 MeV.

SM calculations depend on further assumptions, under which e.g.\ 
$\mu \to \tau$ at $\sqrt{s}= 17.8$ GeV is $\sim 3$
orders smaller than DSM.

\section{Possible experimental tests}
The DSM predictions for the transmutational decay
modes: $\pi^0
\to \mu^- e^+,\ \psi \to \tau^- \mu^+,\ \Upsilon \to \tau^- \mu^+$
could be near experimental limits and sensitivites, for LEP, BEPC and
B-factories. 

For photo-transmutations, one may consider virtual $\gamma$ from $e^+
e^-$ colliders.  Above $\tau$, it is more
profitable 
to look for\cite{study} $e \to \tau$.  Below $\tau$, it is more 
profitable
to look for $e \to \mu$, at around 200 MeV c.m.\ energy.  Again, LEP
and/or BEPC may provide tests.

\section{Conclusions}
\begin{itemize}
\item Lepton transmutation is a necessary consequence in both SM and DSM.
\item The SM results are in general smaller than DSM results, but
there are uncertainties and further assumptions.
\item The DSM calculations are entirely parameter-free.    There are no
violations of data in all the cases we were able to consider.
\item Experimental tests of mainly DSM predictions seem feasible in
the near future, for the following:
\begin{itemize}
\item decays of $\pi^0,\psi,\Upsilon$,
\item photo-transmutation of $e \to \tau$ at high and $e \to
\mu$ at low energy,
\item other processes e.g.\ $e^+ e^- \to e^+ \mu^-$.
\end{itemize}
\item Whether our results will pass these tests or not, there is
exciting new physics that has to be explored.
\end{itemize}

\section*{Acknowledgments}
I thank the Royal Society for a travel grant.


\begin{thebibliography}{99}
\bibitem{expmix} Soudan, Chooz and Superkamiokande data: full
   references in our papers\cite{impromat,photrans}.
\bibitem{bimax} See e.g.\ references in our
   papers\cite{impromat,photrans}.
\end{thebibliography}
\end{document}